\documentclass[reprint,aps,twocolumn,showpacs,nofootinbib]{revtex4-1}

\usepackage{color}

\definecolor{green}{rgb}{0,.5,0}

\definecolor{red}{rgb}{1,0,0}

\usepackage{graphicx}
\usepackage[subfigure]{graphfig}
\usepackage{epsfig}
\usepackage{dcolumn}
\usepackage{amsmath}
\usepackage{rotating}
\usepackage{latexsym}

\DeclareMathOperator{\Tr}{Tr}

\def\be{\begin{equation}}
\def\ee{\end{equation}}
\def\bea{\begin{eqnarray}}
\def\eea{\end{eqnarray}}

\begin{document}

\title{A Demonstration of Hadron Mass Origin from QCD Trace Anomaly}

\author{Fangcheng He$^{1}$, Peng Sun$^{2}$, Yi-Bo Yang$^{1,3,4}$
\vspace*{-0.5cm}
\begin{center}
\large{
\vspace*{0.4cm}
\includegraphics[scale=0.15]{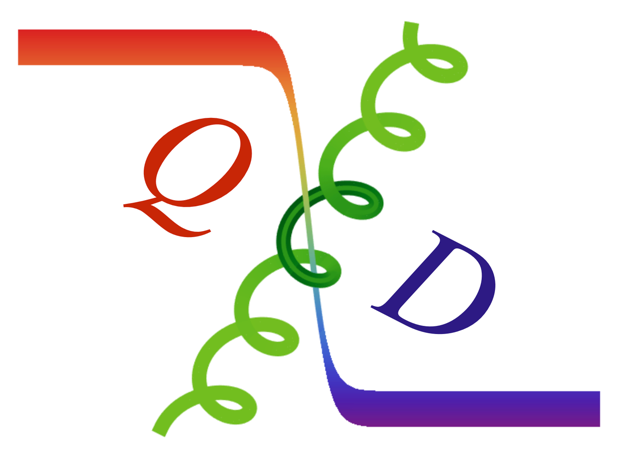}\\
\vspace*{0.4cm}
($\chi$QCD Collaboration)
}
\end{center}
}
\affiliation{
$^{1}$\mbox{CAS Key Laboratory of Theoretical Physics, Institute of Theoretical Physics, Chinese Academy of Sciences, Beijing 100190, China}\\
$^{2}$\mbox{Nanjing Normal University, Nanjing, Jiangsu, 210023, China}\\
$^{3}$\mbox{School of Fundamental Physics and Mathematical Sciences, Hangzhou Institute for Advanced Study, UCAS, Hangzhou 310024, China}\\
$^{4}$\mbox{International Centre for Theoretical Physics Asia-Pacific, Beijing/Hangzhou, China}\\
}

\begin{abstract}
Quantum chromodynamics (QCD) claims that the major source of the nucleon invariant mass is not the Higgs mechanism but the trace anomaly in QCD energy momentum tensor. Although experimental and theoretical results support such conclusion, a direct demonstration is still absent. We present the first Lattice QCD calculation of the quark and gluon trace anomaly contributions to the hadron masses, using the overlap fermion on the 2+1 flavor dynamical Domain wall quark ensemble at $m_{\pi}=340$ MeV and lattice spacing $a=$0.1105 fm. The result shows that the gluon trace anomaly contributes to most of the nucleon mass, and the contribution in the pion state is smaller than that in others nearly by a factor $\sim$10 since the gluon trace anomaly density inside pion is different from the other hadrons and the magnitude is much smaller. The gluon trace anomaly coefficient $\beta/g^3=-0.056(6)$ we obtained is consistent with its regularization independent leading order value $(-11+\frac{2N_f}{3})/(4\pi)^2$ perfectly.
\end{abstract}

\maketitle


\textbf{Introduction:}\ The most essential and non-trivial feature of the quantum field theory, is the regularization. It introduces a tiny and artificial modification on the short distance part of the original theory, to make the contributions from the virtual particle loops to be finite and calculable, while sensitive to the details of the modification. Most of the regularization efforts can be absorbed into the renormalization of fields and the coupling constants, and then irrelevant to the long distance physics. Thus the exceptional cases are named as ``quantum anomaly".

Taking the trace anomaly of the QCD energy momentum tensor (EMT) with quark field $\psi$ and gluon field strength tensor $F_{\mu\nu}$ as an example, the gluon part of the original EMT is traceless, but it is unavoidable to introduce a non-vanishing trace term on the EMT during the regularization, likes the dimensional regularization case discussed in Ref.~\cite{Hatta:2018sqd}. The renormalization eliminates most of the regularization dependence in the additional trace terms and convert them into something proportional to the strong coupling constant, and then the renormalized EMT trace can be expressed as
\begin{align}
T_{\mu}^{\mu}=H_m+(\gamma_mH_m+\frac{\beta}{2g}F^2),
\end{align}
where the quark mass term $H_m=\sum_{q}m_q\bar{q}q$ is the classical trace of EMT, and both the other terms are the trace anomaly. Note that the $F^2$ here is defined in the euclidean space and its vacuum expectation value is positive. The anomalous terms are proportional to the anomalous dimension of the quark mass $m_q$, $\gamma_m=-\frac{\mu}{m}\frac{\partial m}{\partial \mu}=\frac{2}{\pi}\alpha_s+{\cal O}(\alpha_s^2)$, and also that of the strong coupling $\alpha_s=\frac{g^2}{4\pi}$, $\frac{\beta}{2g}=\frac{\mu^2}{2\alpha_s}\frac{\partial \alpha_s}{\partial \mu^2}=(-\frac{11}{8\pi}+\frac{N_f}{12\pi})\alpha_s+{\cal O}(\alpha_s^2)$, respectively.

The trace anomaly leads to the most non-trivial feature of QCD: the quantum particles likes nucleon can have positive masses, even though the classical waves of quark and gluon travel at the speed of light. The argument is simple: the hadron mass is proportional to its matrix element of the EMT trace~\cite{Shifman:1978zn,Ji:1994av,Ji:2021pys},
\begin{align}\label{eq:sum_rule}
M_{H}=\langle T_{\mu}^{\mu} \rangle_H = (1+\gamma_m)\langle H_m\rangle_H+\frac{\beta}{2g} \langle F^2\rangle_H,
\end{align}
with $\langle O\rangle_H\equiv \langle H| O|H\rangle$; thus the gluon trace anomaly $\frac{\beta}{2g} \langle F^2\rangle_H$ can contribute a positive hadron mass even in the chiral limit  where $m_q\rightarrow 0$ and $\langle H_m\rangle_H$ vanish. The only exception is pion: its mass will be zero in the chiral limit and then the percentage of trace anomaly contribution in pion mass should be less than 50\%~\cite{Ji:1995sv}. One can also use this picture to understand the QED trace anomaly effect in the lamb shift, 
while it turns out to be small~\cite{Sun:2020ksc,Ji:2021pys}.

The above argument can be verified indirectly through the sum rule in Eq.~(\ref{eq:sum_rule}): The nucleon mass $M_N$ can be measured at $10^{-10}$ accuracy, the $\langle O\rangle_H$ from three light flavors can be extracted from the SU(3) flavor breaking of the baryons and/or explicit Lattice QCD calculations. Then their difference ($\sim$ 850 MeV~\cite{Yang:2015uis,Yang:2018nqn}) should come majorly from the trace anomaly. The asymptotic freedom and confinement of QCD require the QCD coupling to be stronger at larger distance and then $\beta$ is negative, thus $\langle F^2\rangle_H$ should also be negative to satisfy the sum rule, even though the matrix element $\langle F^2\rangle$ in the vacuum will be positive. Note that the heavy quark contribution is canceled by the flavor dependence of $\frac{\beta}{2g}$ at leading order based on the heavy quark approximation~\cite{Shifman:1978zn,Tarrach:1981bi},
\begin{align}\label{eq:heavy_quark}
m_Q\bar{Q}Q\  _{\overrightarrow{m_Q\rightarrow \infty}}\ -\frac{\alpha_s}{12\pi}F^2+{\cal O}(\alpha_s^3),
\end{align}
and then is effectively decoupled.

The experimental measurement of trace anomaly arouses great interest, and is considered as one of the major scientific goals of the future Electron-Ion Collider (EIC)~\cite{Accardi:2012qut} and EicC~\cite{Anderle:2021wcy}. One approach to detect trace anomaly is measuring the cross section of exclusive photo-production of heavy quarkonium which depends on the gluon condensate at non-relativistic limit~\cite{Luke:1992tm,Kharzeev:1995ij,Kharzeev:1998bz}. The gluon gravitational form factors which can relate to $J/\Psi$ production amplitude in the large momentum limit, is also a possibility worth further investigation~\cite{Boussarie:2020vmu}. There are also other theoretical discussion on the related experimental possibility~\cite{Sun:2021gmi,Guo:2021ibg}. 

In theoretical side, the central challenge is verifying whether the difference between hadron mass and the its quark mass contribution actually comes from the quantum anomaly effort. 
Such a calculation is highly non-perturbative and then can only be done with Lattice QCD, but the chiral symmetry breaking in the quark mass term of the Wilson-like action can mixed with the original trace anomaly. Thus the expensive chiral fermion is essential
and there is no explicit verification yet.

In this work, we present the first demonstration of the quantum trace anomaly contribution to kinds of the hadron masses, after this mass generation mechanism has been proposed for more than 40 years. We confirm that Eq.~(\ref{eq:sum_rule}) is satisfied in all hadrons we calculated in this work with the $\gamma_m$ and $\beta$ determined non-perturbatively. In addition, we also find that the gluon trace anomaly density in the pion turns out to be much smaller than that in the other hadron likes nucleon and vector meson, due to the significant difference on the gluon trace anomaly distribution inside the hadrons.

\begin{table}[ht!]
\caption{Information of the RBC ensemble~\cite{Blum:2014tka} used in calculation. The pion and kaon masses are in unit of MeV.}
\begin{ruledtabular}
\begin{tabular}{l l l c c l c  c}
\text{Symbol} & $L^3 \times T$  &  $a$ (fm)  & $6/g^2$  & $m_\pi$ & $m_K$ &  $N_\text{cfg}$\\
\hline
24I  & $24^3\times\ 64$& 0.1105(3) & 2.13 &340 & 593 &203  \\
\end{tabular}
\end{ruledtabular}
\label{table:ensemble}
\end{table}

\textbf{Numerical setup}: We preform the calculation on a $24^3\times 64$ 2+1 flavor Domain-wall fermion ensemble from the RBC collaboration~\cite{Blum:2014tka}, with the pion mass $m_{\pi}=340$ MeV which is heavier than the physical one but still much smaller than the corresponding nucleon mass. The other information of the ensemble we use summarized in Table~\ref{table:ensemble}. For the valence quark, we use the chiral fermion through the overlap approach~\cite{Chiu:1998gp} to avoid the additional term in the trace of EMT under the lattice regularization.

In order to determine the $\gamma_m$ and $\beta$ precisely and make an accurate test of the sum rule, we consider the partially quenched QCD~\cite{Bernard:1993sv} which allows the valence quark mass to be different from that in the gauge ensemble.
In such a case, Eq.~(\ref{eq:sum_rule}) becomes
\begin{align}\label{eq:sum_rule2}
M_{H}= (1+\gamma_m)\big(\langle H_m\rangle^v_H+\langle \sum_im_i\bar{q}_iq_i\rangle_H\big)+ \langle H^g_a\rangle_H,
\end{align}
where $H^g_a=\frac{\beta}{2g}F^2$, $\langle H_m\rangle^v_H$ includes the connected quark diagram with the operator $m_v\bar{q}_v q_v$ only,
and the index $i$ just includes the flavors exist in the gauge configurations: degenerated light up/down and also strange quark. 

In this work, we considered the cases with 5 quark masses $m_va$=0.0160, 0.0576, 0.1, 0.2, and 0.3, and the lightest two quark masses correspond to the light and strange quark masses in the gauge ensemble we used. 

First of all, the hadron mass can be extracted from the two point correlation function with wall source and wall sink
\begin{align}
C_2(t_f;\cal H)&=\langle \sum_{\vec{y}} {\cal H}(t_f,\vec{y}) \sum_{\vec{x}}{\cal H}^{\dagger}(0,\vec{x})\rangle,
\end{align}
where ${\cal H}$ is the hadron interpolation field.
As implemented in Ref.~\cite{Sun:2020pda}, when we calculate the contribution of valence quark mass, we use the Feynman–Hellmann theorem inspired method~\cite{Chang:2018uxx} to construct the summed three point for the connected insertion case, 
\begin{align}\label{eq:3ptc}
SC^{q_v}_3(t_f;\cal H)&=\langle \sum_{\vec{y}} {\cal H}(t_f,\vec{y})\sum_{t,\vec{z}}O^{q_v}(t,\vec{z})\sum_{\vec{x}}{\cal H}^{\dagger}(0,\vec{x})\rangle,\nonumber\\
\end{align}
where $O^{q_v}=m_q^v\bar{q}_vq_v$ is the valence quark operator. For the disconnected insertion case, we modified the expression into 

\begin{align}\label{eq:3pt}
&SC^{q_s,g}_3(t_f;{\cal H})=\nonumber\\
&\quad \langle \sum_{\vec{y}} {{\cal H}}(t_f,\vec{y}) \!\!\! \sum_{t\in(t_f,0),\vec{z}}\!\! O^{q_s,g}(t,\vec{z}) \sum_{\vec{x}}{\cal H}^{\dagger}(0,\vec{x})\rangle,
\end{align}
where the sum of the current time slides in the light sea quark operator $O_q=m_s\bar{q}_s{q}_s$ ($m_s$ is sea quark mass) and gluon operator $O^g=F^2$ are limited to $t\in(t_f,0)$ to remove the statistical uncertainty from the unphysical region $t<0$ and $t>t_f$, and the gluon field tensor $F_{\mu\nu}$ is defined through the standard clover definition.
The detailed expression of $C_2$ and $SC_3$ and how to construct them with propagators, are shown in the supplemental materials~\cite{sm_anomaly}. 

For each hadron, we carry out a joint correlated fit of  $C_2(t_f;H)$  and $SC^{q,g}_3(t_f;H)$ to extract the $M_H$, $\langle H_m\rangle_H$ and $\langle F^2 \rangle_H$ simultaneously. As mentioned above, $SC^{q_v,q_s,g}_3(t_f;H)$ is summed three point correlation function which has summed the total contribution of matrix element of the interpolated quark and gluon operator between 0 and $t_f$. The fit expression can be written as ~\cite{Sun:2020pda}:
\begin{align}\label{eq:fit}
SC^{i=q_v,q_s,g}_3(t_f;H)=&e^{-M_Ht_f}\left(B_0 t_f \langle O^{i}\rangle_H + B^{i}_{2}e^{-\delta m t_f} \right.\nonumber\\
&\left.+ B^{i}_3t_f e^{-\delta m t_f}+B^{i}_4\right)\nonumber\\
C_2(t_f;H)=&B_0e^{-M_H t_f} (1+B_1e^{-\delta m t_f}),
\end{align}
where $\langle O^{q_v,q_s}\rangle_H=\langle H^{v,s}_m\rangle_H$, $\langle O^g\rangle_H=\langle F^2\rangle_H$, the $e^{-\delta m t_f}$ terms are introduced to account for the contamination from higher states, $\delta m$, $B_0$, $B_1$ and $B^{q_v,q_s,g}_{2,...,4}$ are free parameters. The contribution from transition between ground state and higher excited state are included in $B^{q_v,q_s,g}_{2,...,4}$. The above form is equivalent to extract the desired quantities in the large $t_f$ limits~\cite{Sun:2020pda,Chang:2018uxx},
\begin{align}\label{eq:matrix_element}
M_H&=\textrm{log}\frac{C_2(t_f-1;H)}{C_2(t_f;H)}+{\cal O}(e^{-\delta m t_f}),\nonumber\\
 \langle H_m\rangle_H&=\Delta R^q(t_f;H)+{\cal O}(e^{-\delta m t_f}),\nonumber\\
  \langle F^2\rangle_H&=\Delta R^g(t_f;H)+{\cal O}(e^{-\delta m t_f}),
\end{align}
where $ \Delta R^{i={q,g}}(t_f)\equiv \frac{SC^i_3(t_f;H)}{C_2(t_f;H)}-\frac{SC^i_3(t_f-1;H)}{C_2(t_f-1;H)}$. Note that one need to replace $SC_{2,3}(t_f)$ into $SC_{2,3}(t_f)+SC_{2,3}(Ta-t_f)$ for the PS meson case, to include the loop around effect and describe the data around $t_f\sim Ta/2$.

\textbf{Result:}\
In this work, we calculated the quark propagator at five different valence quark masses $m_va$=0.0160, 0.0576, 0.1, 0.2, and 0.3 respectively, and construct the two and three point correlation functions for the nucleon (N), pseudoscalar (PS), and vector (V) mesons. We also looped over all the $T=64$ time slides to suppress the statistical uncertainty, and applied 5 steps of the HYP smearing on the gauge operator. 

\begin{figure}[htbp]
	\centering
	\includegraphics[width=0.4\textwidth]{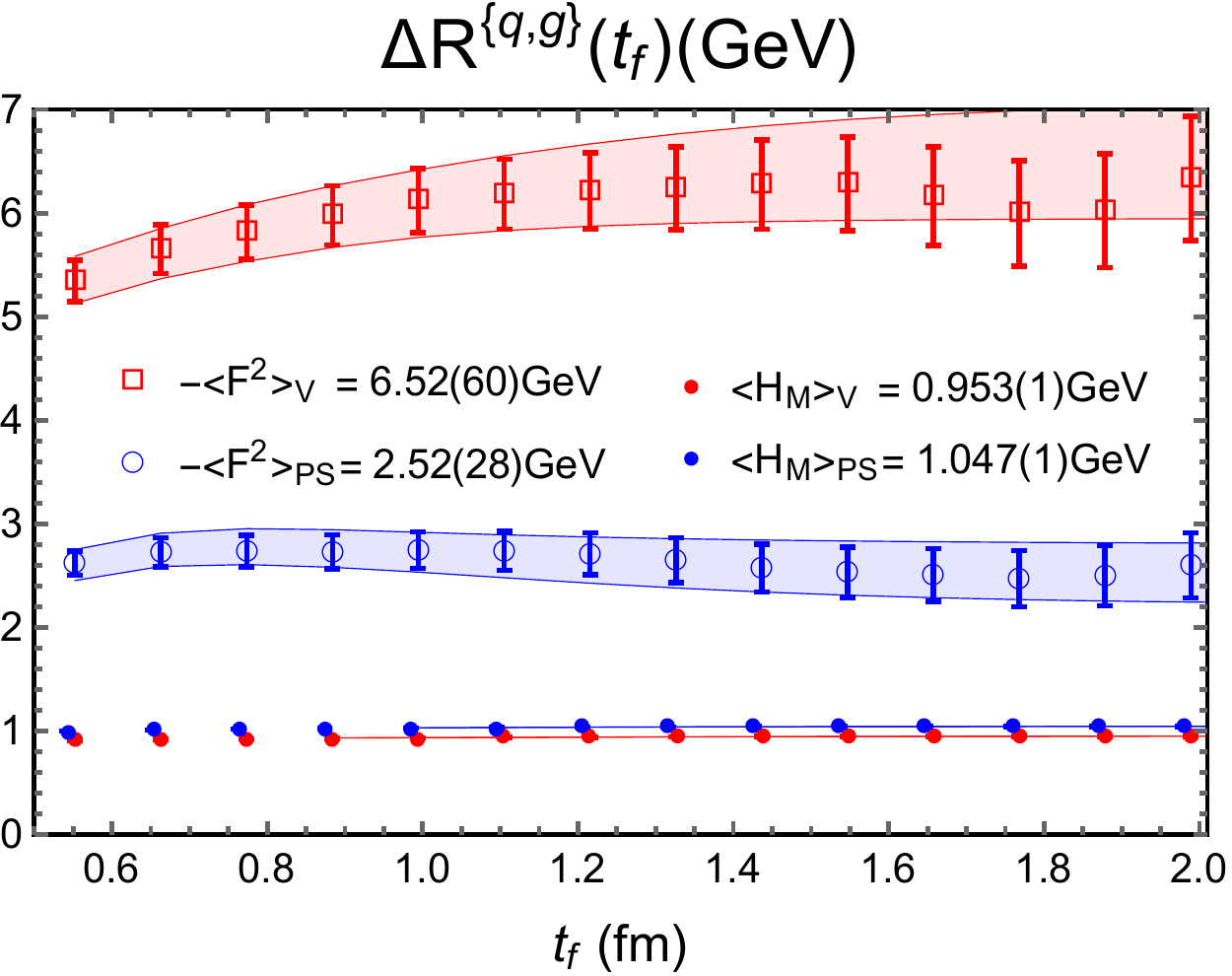}
	\caption{The differential ratio $\Delta R^{q,g}_{PS, V}(t_f)$ of the pseudoscalar and vector mesons with $m_va=0.3$, which should approaches to the ground state matrix elements $\langle H_m\rangle_{PS, V}$ and $-\langle F^2\rangle_{PS, V}$ respectively at the $t_f\rightarrow\infty$ limit. The bands shows the joint fit prediction using the form defined in Eq.~(\ref{eq:fit}) and they agree with the data well.}
	\label{fig:plateau}
\end{figure}

In Fig.~\ref{fig:plateau}, we plotted the effective ratios $\Delta R^{q,g}(t_f)$ defined in Eq.~(\ref{eq:matrix_element}) for the PS (blue) and V (red) mesons with $m_va$=0.3. The disconnected light quark contributions are not shown here as their contributions are small. The $M_H$, $\langle H_m\rangle_H$ and $\langle F^2 \rangle_H$ are obtained using the joint fit defined in Eq.~\ref{eq:fit} with $\chi^2$/d.o.f.$\sim$1, and and the bands of $\Delta R^{q,g}(t_f)$ predicted by the joint fit are also shown in Fig.~\ref{fig:plateau} and agree with the data perfectly. We can see that the gluon trace anomaly matrix element $-\langle F^2\rangle_H=-\Delta R^g(t_f)|_{t_f\rightarrow \infty}$ in the V meson is more than two times of that in the PS meson, even though their quark mass terms $\langle H_m\rangle_H=\Delta R^q(t_f)|_{t_f\rightarrow \infty}$ just differ from each other by 10\%.
The values of $M_H$, $\delta_m$, excited state mass difference $\delta m$, fitted matrix elements and $\chi^2$/d.o.f. of different hadrons with kinds of quark masses can be founded in the supplemental materials~\cite{sm_anomaly}.

Since the anomalous dimension $\gamma_m$ and $\beta$ should be independent to the hadron states, we solve the equations
\begin{align}\label{eq:sol}
&M_{PS}- (1+\gamma_m)\langle H_m\rangle_{PS}-\frac{\beta}{2g} \langle F^2\rangle_{PS}|_{m_va=0.3}=0,\nonumber\\
&M_{V}- (1+\gamma_m)\langle H_m\rangle_{V}-\frac{\beta}{2g} \langle F^2\rangle_{V}|_{m_va=0.3}=0,
\end{align}
and obtain the bare $\gamma_m=0.38(3)$ and $\frac{\beta}{2g}=-0.08(1)$. Both $\gamma_m$ and $\frac{\beta}{2g}$ are significantly smaller than one due to the suppression of the effective coupling constant $\alpha_s$, while the $\langle F^2\rangle_{PS/V}$ is large enough to reverse the naive power counting. At the same time, since $\langle g^2F^2\rangle_{H}$ is the function of the gauge link without the bare coupling dependence explicitly, it is more natural to divide $g^2$ on the coefficient $\frac{\beta}{2g}$ and consider $\frac{\beta}{g^3}=-0.056(6)$ which is independent of $\alpha_s$ at the leading order. Such a value is perfectly agree with the regularization independent leading order $\frac{\beta^{(0)}}{g^3}=(-11+\frac{2N_f}{3})/(4\pi)^2=-0.057(7)$ with the uncertainty from the next to leading order correction using the bare $\alpha_s$. In the other hand, the $\gamma_m=0.38(3)$ we obtained is comparable with that under the $\overline{\textrm{MS}}$ ($1/a$=1.78 GeV) which is 0.325(10) with the error estimated from the ${\cal O}(\alpha_s^4)$ correction.

\begin{figure}[htbp]
	\centering
	\includegraphics[width=0.48\textwidth]{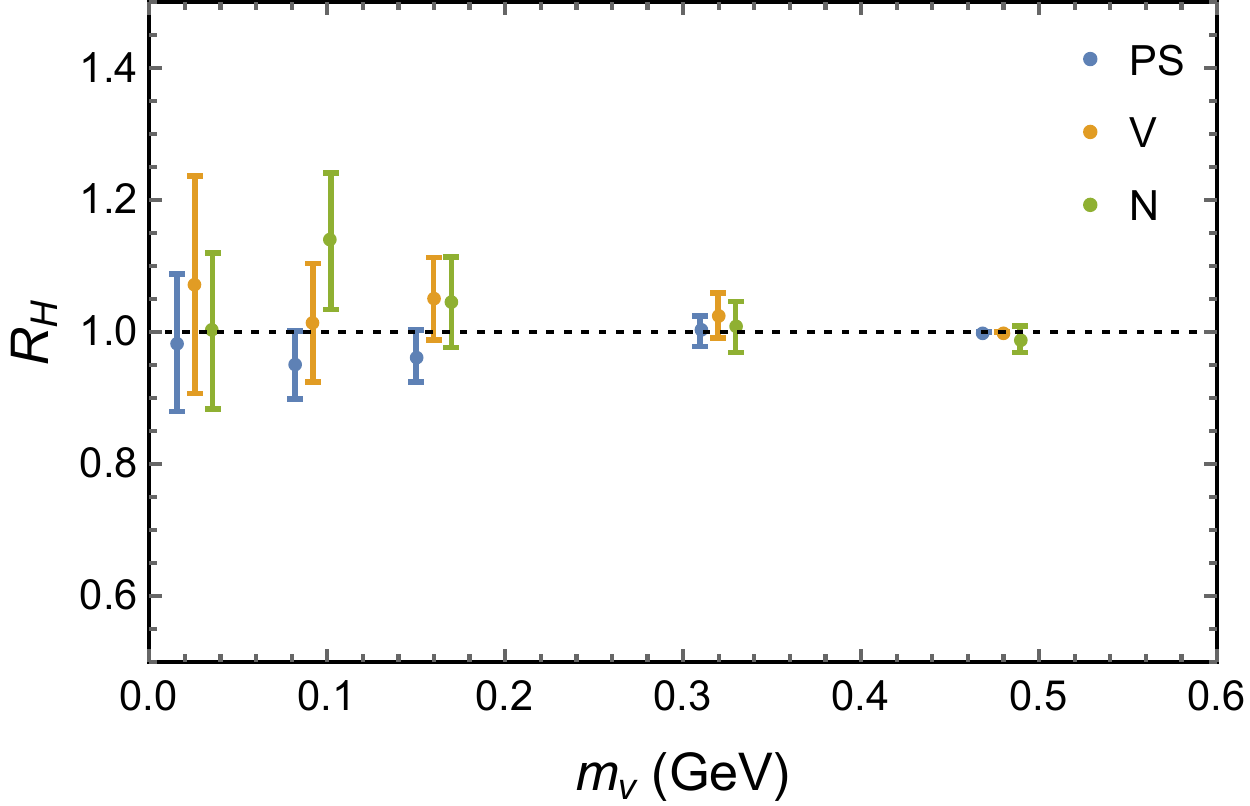}
	\caption{The ratio of the hadron mass from the left and right hand side of Eq.~(\ref{eq:sum_rule}), using the $\gamma_m$ and $\beta/g$ obtained from the PS and V mesons with $m_va$=0.3. It is consistent with one within the uncertainties in all the cases.}
	\label{fig:ratio}
\end{figure}

With the above values of $\gamma_m$ and $\frac{\beta}{2g}$, we calculate the hadron mass $M_H$, quark mass term $\langle H_m\rangle_{H}$ and $\langle F^2\rangle_{H}$ of the pseudoscalar, vector meson and nucleon with different $m_v$, and plot the ratio,
\begin{align}
R_H(m_v)=\frac{(1+\gamma_m)\langle H_m\rangle_{H}+\frac{\beta}{2g} \langle F^2\rangle_{H}}{M_H},
\end{align}
in Fig.~\ref{fig:ratio}. We can see that the $R_H$ in all the cases are consistent with one within the uncertainties. Note that the PS and V meson cases with $m_va=0.3$ are exactly one since they are the input cases. Such a result verifies the trace anomaly sum rule in Eq.~(\ref{eq:sum_rule}), and is consistent with our expectation that both $\gamma_m$ and $\beta$ are universal.

\begin{figure}[htbp]
	\centering
	\includegraphics[width=0.48\textwidth]{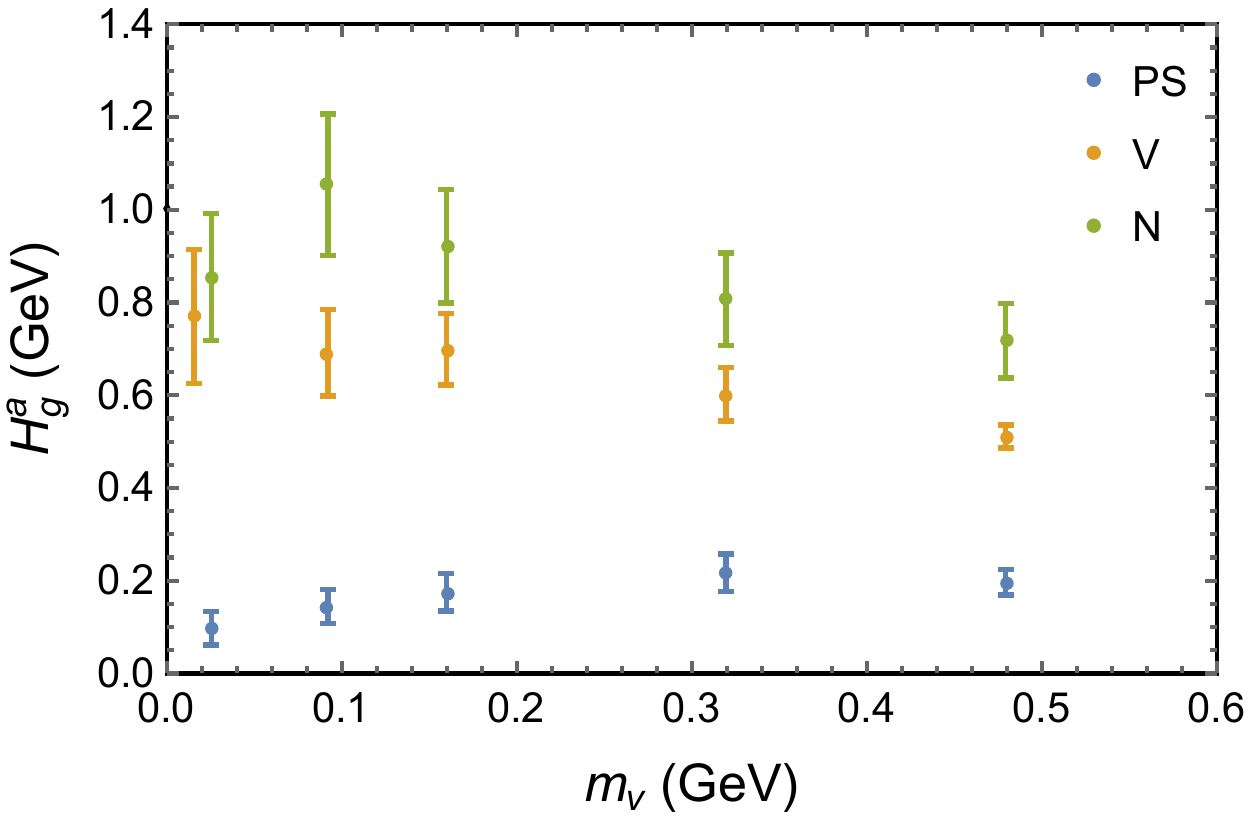}
	\caption{The gluon trace anomaly contribution to the hadron mass. We can see that it is is always small in the PS meson, while approaches to $\sim$ 800 MeV for the nucleon and vector meson in the chiral limit $m_v\rightarrow 0$.}
	\label{fig:gluon}
\end{figure}

The resulting gluonic trace anomaly contribution $\langle H_a^g \rangle_H=\langle \frac{\beta}{2g}F^2 \rangle_H$ in the 2+1 flavor ensemble are plotted in Fig.~\ref{fig:gluon}. We can see that $\langle H_a^g \rangle_H$ in the pion state is generally smaller than that in the other states, especially at the unitary point with 340 MeV pion mass. In such a case, the gluon trace anomaly contributes about 100 MeV of the pion mass which is $\sim$30\%, but $\sim$ 800 MeV in the $\rho$ meson and also nucleon. It is exactly what we expected from the trace anomaly sum rule: The trace anomaly contributes most of the hadron masses, except the pion case.

\begin{figure}[htbp]
	\centering
	\includegraphics[width=0.48\textwidth]{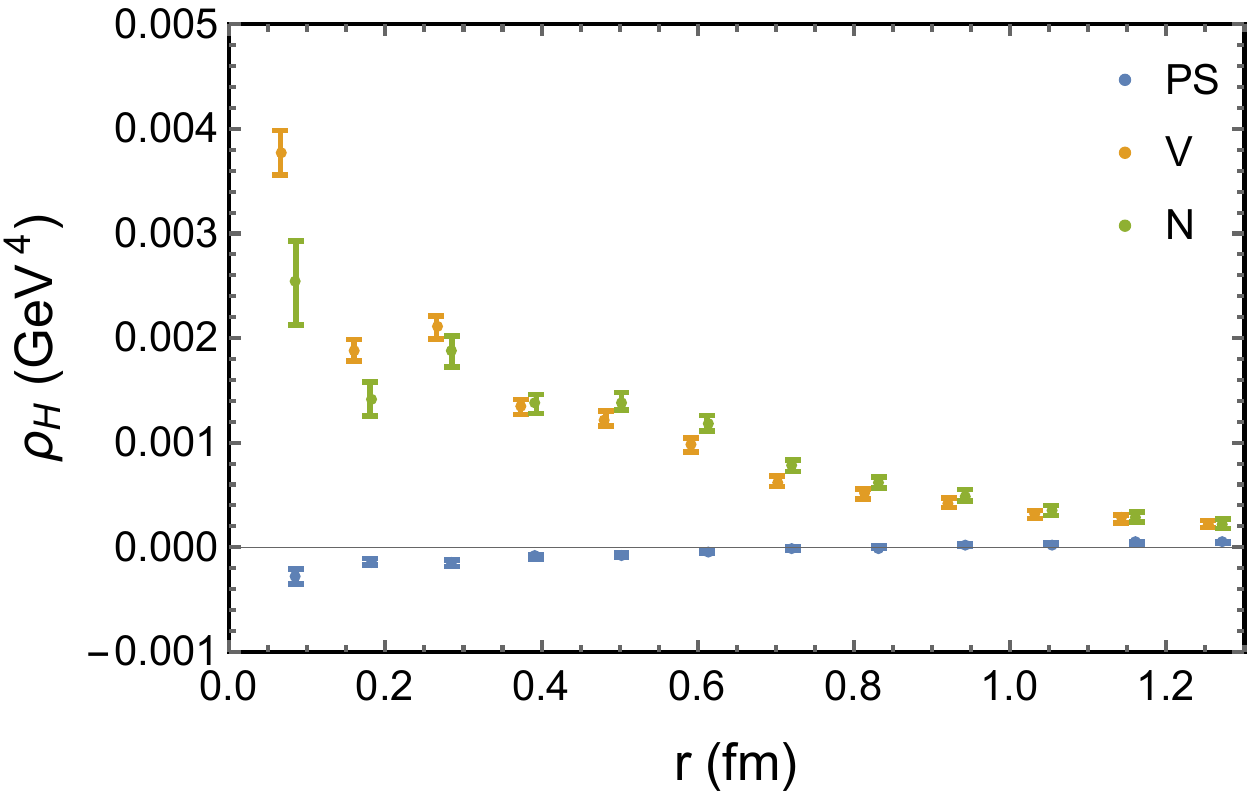}
	\caption{The density of gluon trace anomaly in the different hadron state with 340 MeV pion mass.}
	\label{fig:density}
\end{figure}

The difference will be much more significant at the physical quark mass. If we use the GMOR relation $m_{\pi}^2\propto m_q$ and the Feynman-Hellman theorem $\langle H_m\rangle_{H}=m_q\frac{\partial m_{\pi}}{\partial m_q}=\frac{1}{2}m_{\pi}+{\cal O}(m_{\pi}^3)$~\cite{Gasser:1982ap,Ji:1995sv}, we can estimate the gluon trace anomaly contribution in the physical pion state to be $\frac{1}{2}(1-\gamma_m)m_{\pi}=43(2)$ MeV. On the other hand, that in the nucleon at the physical light quark mass will be 816(10) MeV if we just consider the quark mass contribution from three light quarks~\cite{Yang:2015uis,Yang:2018nqn}.

In order to uncover the origin of this difference, we also investigate the gluon trace anomaly density inside the hadron,
\begin{align}
&\rho_{H}(|r|)=\nonumber\\
&\frac{\langle \sum_{\vec{y}} {\cal H}(t_f,\vec{y})H_a^g(t,\vec{y}+\vec{r})\sum_{\vec{x}}{\cal H}^{\dagger}(0,\vec{x})\rangle}{\langle \sum_{\vec{y}} {\cal H}(t_f,\vec{y})\sum_{\vec{x}}{\cal H}^{\dagger}(0,\vec{x})\rangle}|_{t,t_f-t\rightarrow \infty},
\end{align}
as $\rho_{H}(|r|)$ can be related to the gluon trace anomaly matrix element $\langle H_a^g\rangle_H=\int_{\vec{r}}\textrm{d}^3r \rho_{H}(|r|)$ and its squared charge radius $\langle r^2_g\rangle_{H}=\frac{\int_{\vec{r}}\textrm{d}^3r r^2 \rho_{H}(|r|)}{\langle H_a^g\rangle_H}$, with proper integration like the pion change radius case studied in Ref.~\cite{Feng:2019geu}. A naive guess is that $\rho_{H}$ would be insensitive to the hadron at small $r$, while it should decay much faster in the pion than the other hadrons to make the integration and change radius to be significantly smaller. But our lattice QCD calculation has excluded such a possibility, as shown in Fig.~\ref{fig:density}. First of all, $\rho_H$ is much smaller than the leading order estimate of the vacuum expectation value $\langle F^2\rangle=0.012(4)$~\cite{Shifman:1978by}, even at small $r$. At the same time, one can see that the density of pion is much smaller than that of nucleon and vector meson; even more, the density tends to be negative and then have the same sign as that in the vacuum.  This means that $\langle H_a^g\rangle$ receives both suppression from the magnitude of density and also short range interaction, and it can make the pion gluon trace anomaly radius to be larger than the other hadrons. More details of distribution calculation can be found in the supplemental materials~\cite{sm_anomaly}.


\textbf{Summary and outlook}:\
In this work, we calculated the quark mass term $\langle m\bar{\psi}\psi\rangle_{H}$ and gluon action terms $\langle F^2\rangle_{H}$ in the hadron. Based on the EMT trace anomaly sum rule in the PS and V meson states with $m_va$=0.3, we determined the bare anomalous dimensions of the quark mass and gluon coupling constant as $\gamma_m=0.38(3)$ and $\frac{\beta}{g^3}=-0.056(6)$ (with 5 steps of HYP smearing on the gluon operator) respectively, and confirmed that they are independent on hadron states and quark mass up to the statistical uncertainties. With such $\gamma_m$ and $\frac{\beta}{g^3}$, we find that the gluon trace anomaly contribution in the PS meson mass is always much smaller than that in the other hadrons, especially around the chiral limit. 

A more accurate check on the trace anomaly mechanism can be carried out by renormalizing the $\langle F^2\rangle_H$ non-perturbatively~\cite{Yang:2018bft} and convert it to that under the $\overline{\textrm{MS}}$ scheme, and/or deducing the conserved EMT with ${\cal O}(a^2)$ trace term directly from the discretized QCD action. The calculation with different gauge actions with ${\cal O}(a^2)$ difference on the bare $g^2$ at the same lattice spacing (likes Symanzik action used by the MILC configurations and Iwasaki action used here), and the lattice spacing dependence is also desirable. Besides, a direct calculation of the heavy quark contribution $\langle m_Q\bar{\psi}_Q\psi_Q\rangle_H$ can also provides a non-varification through Eq.~(\ref{eq:heavy_quark}). 

A non-perturbative determination of $\gamma_m$ and $\frac{\beta}{g^3}$ at given lattice spacing opens kinds of the interesting possibility on further studies. First of all, Currently the QCD equation of state (EOS) at finite temperature uses either the lattice spacing dependence of the coupling constants~\cite{Bazavov:2014pvz}, or the gradient flow~\cite{Carosso:2018bmz} to determine $\gamma_m$ and $\frac{\beta}{g^3}$, and then Eq.~\ref{eq:sum_rule} can only be hold after the continuum extrapolation. The scheme we demonstrated here allows an accurate determination of EOS at any given lattice spacing and then suppresses the discretization errors. It also allows us to calculate the proton gravity form factor and radius directly at the physical pion mass but not only the chiral limit~\cite{Kharzeev:2021qkd}. At the same time, it suggests that the coupling between hadrons and the bi-linear heavy quark operator will be very case sensitive based on the relation Eq.~(\ref{eq:heavy_quark}). For example, the total trace anomaly contribution including both the $\gamma_m$ and $\frac{\beta}{2g}$ terms are generally $\sim$ 800 MeV for most of hadrons except light pseudoscalar mesons. But with $\gamma_m\sim 0.3$, we will have $\langle F^2\rangle_{N}\sim 0.84$ GeV and $\langle F^2\rangle_{\eta_c}\sim 0.12$ GeV. And then $\langle m_b\bar{\psi}_b\psi_b\rangle_{N}$ will be larger than $\langle m_b\bar{\psi}_b\psi_b\rangle_{\eta_c}$ by a factor of $\sim$ 5 and should be verified with direct calculations. Another possibility is related to the quark and gluon ${\cal O}(\Lambda_{QCD})$  ``dynamical mass'' at small off-shell region~\cite{Bowman:2005vx,Bicudo:2015rma} under the Landau gauge, whose origin and the relation with the hadron mass are still unclear yet. In term of the EMT matrix elements in the parton state, the gauge independent trace anomaly and/or the gauge dependent ones should be the origin of such a mass. Thus practical calculation of the trace anomaly matrix element using the coefficient $\frac{\beta}{g^3}$ obtained here can unravel the role of quantum anomaly efforts in the dynamically generated parton masses.


Another interesting thing is the gluon trace anomaly densities in the pion and nucleon are comparable in their extent, while their magnitude are very different. Even more, the density in the center of pion is negative, and then have the same sign as that in the vacuum. It open a new gate to understand the relation between pion and vacuum, and why the other hadrons are different from them.


\section*{Acknowledgement}
We thank Heng-Tong Ding, Xiangdong Ji, Luchang Jin, Keh-Fei Liu, Jianhui Zhang, and Jian Zhou for valuable discussions, and the RBC and UKQCD collaborations for providing us their DWF gauge configurations. The calculations were performed using the GWU-code~\cite{Alexandru:2011ee,Alexandru:2011sc} through HIP programming model~\cite{Bi:2020wpt}. The numerical calculation has majorly been done on CAS Xiaodao-1 computing environment,  and supported by Strategic Priority Research Program of Chinese Academy of Sciences, Grant No. XDC01040100, and also HPC Cluster of ITP-CAS, Jiangsu Key Lab for NSLSCS.
P. Sun is supported by Natural Science Foundation of China under grant No. 11975127, as well as Jiangsu Specially Appointed Professor Program. Y. Yang is  supported by Strategic Priority Research Program of Chinese Academy of Sciences, Grant No. XDB34030303, XDC01040100 and XDPB15. P. Sun and Y. Yang are also suppored in part by a NSFC-DFG joint grant under grant No. 12061131006 and SCHA~458/22.

\bibliographystyle{apsrev4-1}
\bibliography{reference.bib}

\begin{widetext}

\section*{Supplementary materials}

\subsection{Contraction of the Green functions}
In this section, we will give the expressions of two point and summed three point correlation functions constructed by propagator. In our calculation, we use the coulomb gauge fixed wall source propagator $S_w(\vec{y},t_2; t_1)=\sum_{\vec{x}}S(\vec{y},t_2; \vec{x},t_1)$ and the Feynman–Hellmann propagator~\cite{Chang:2018uxx} $\tilde{S}_{c}(\vec{y},t_2;t_1)=m_q\sum_{\vec{x},t}S(\vec{y},t_2;\vec{x},t)S_w(\vec{x},t;t_1)$ to construct the two point and connected three point correlation functions,
\begin{align}
&C_2(t_f;{\cal M})=\sum_{\vec{y}}C_2(t_f,\vec{y};{\cal M})=C_{\cal M}(S_w,S_w,t_f,0),\nonumber\\
&SC^{q_v}_3(t_f;{\cal M})=C_{\cal M}(S_c,S_w,t_f,0)+C_{\cal M}(S_w,S_c,t_f,0),
\end{align}
where $C_{\cal M}$ are defined as
\begin{align}
&C_{\cal M}(S_1,S_2,t_f,0)=\sum_{\vec{y}}\langle \textrm{Tr}\left(\gamma_5S^{\dagger}_1( \vec{y},t_f;0)\gamma_5 \Gamma S_2(\vec{y},t_f; 0) \Gamma \right)\rangle,
\end{align}
For the meson ${\cal M}(\Gamma)$ with the interpolation field 
$\bar{\psi}\Gamma\psi$, where ${\cal I}$ is the unitary matrix, and
$S(\vec{y},t_2; \vec{x},t_1)$ is the quark propagator from $(\vec{x},t_1)$ to
$(\vec{y},t_2)$. For quark loop and gluon operator, we need calculate the
following disconnected three point correlation function,
\begin{align}\label{eq:totm}
SC_3^{q_s,g}(t_f,{\cal M})
=\sum_{0<t<t_f}\sum_{\vec{y}}\sum_{\vec{x}}\Big\langle [C_2(t_f,\vec{y};{\cal M})-\langle C_2(t_f,\vec{y};{\cal M})\rangle][{\cal O}^{q_s,g}(t,\vec{x})-\langle {\cal O}^{q_s,g}(t,\vec{x})\rangle]\Big\rangle
\end{align}
The definition of ${\cal O}^{q_s}$ and ${\cal O}^{g}$ is in the following content of Eq.~(\ref{eq:3pt}). The nucleon case with the interpolation filed $(u^{T}\tilde{C}d)u$ is similar and can be obtained without much modifications,
\begin{align}\label{eq:3pt_N}
&C_2(t_f;{\cal N})=\sum_{\vec{y}}C_2(t_f,\vec{y};{\cal N})=C_{\cal N}(S_w,S_w,S_w,t_f,0),\nonumber\\
&SC^{q_v}_3(t_f;{\cal N})=C_{\cal N}(S_c,S_w,S_w,t_f,0)+C_{\cal N}(S_w,S_c,S_w,t_f,0)+C_{\cal N}(S_w,S_w,S_c,t_f,0),\nonumber\\
&C_{\cal N}(S_1,S_2,S_3,t_2,t_1)=\langle \epsilon^{abc} \epsilon^{a^\prime b^\prime c^\prime} \sum_{\vec{y}}\Tr\left( \Gamma_m S_1^{aa^\prime}( \vec{y},t_f;0) \right) \Tr\left( \underline{S}^{bb^\prime}_2(( \vec{y},t_f;0) S^{cc^\prime}_3( \vec{y},t_f;0)\right)\nonumber\\
&+\Tr\left( \Gamma_m S^{aa^\prime}_1( \vec{y},t_f;0) \underline{S}^{bb^\prime}_2( \vec{y},t_f;0)S^{cc^\prime}_3( \vec{y},t_f;0)\right)\rangle,
\end{align}
where $\underline{S}$ is defined by $(\tilde{C}S\tilde{C}^{-1})^T$, $\tilde{C} =\gamma_2 \gamma_4 \gamma_5$, and $\Gamma_m=\frac{1}{2}(1+\gamma_4)$ is the unpolarized projector. The summed three point correlation functions for light quark loop and gluon are defined as 
\begin{equation}\label{eq:totn}
SC^{q_s,g}_3(t_f;{\cal N})=\sum_{0<t<t_f}\sum_{\vec{y}}\sum_{\vec{x}}\Big\langle [C_2(t_f,\vec{y};{\cal N})-\langle C_2(t_f,\vec{y};{\cal N})\rangle][{\cal O}^{q_s,g}(t,\vec{x})-\langle {\cal O}^{q_s,g}(t,\vec{x})\rangle]\Big\rangle
\end{equation}

Note that during the calculation with the overlap fermion, deflating the long-distance subspace of the Dirac operator using its eigenvectors $v(\lambda)$ with $|\lambda|<\Lambda_{QCD}$ are essential to obtain the light quark propagator efficiently~\cite{Li:2010pw}. At the same time, we can build the light quark loop operator $O_q(t)$ only via those eigenvectors $v(\lambda)$ with little cost, and the systematic uncertainty from the rest is negligible~\cite{Yang:2015uis},
\begin{align}
O_q(t)&\equiv\sum_{\vec{x}} S(m_q;\vec{x},t;\vec{x},t)\simeq \sum_{\vec{x}} S^L(m_q;\vec{x},t;\vec{x},t), \nonumber\\
S^L(m_q)&\equiv\int^{i\Lambda_{QCD}}_{-i\Lambda_{QCD}} \textrm{d}\lambda \frac{v(\lambda)v^{\dagger}(\lambda)}{\lambda+m_q}.
\end{align}
In practical, 300 pairs of the low lying eigenpairs of the Dirac operator with $\lambda \le 0.246$ GeV are solved to speed up the propagator calculation and construct the light quark loops.

Besides, the clover definition of $F_{\mu\nu}$ used for the gluon operator $F^2$ is the following,
\begin{align}
F_{\mu\nu}(x) &= \frac{i}{8a^2g} \left[\mathcal{P}_{[\mu,\nu]}+\mathcal{P}_{[\nu,-\mu]}+ \mathcal{P}_{[-\mu,-\nu]} + \mathcal{P}_{[-\nu,\mu]} \right](x), \label{eq:emt_clover}\nonumber\\
\mathcal{P}_{\mu,\nu}(x)&= U_\mu(x)U_\nu(x+a\hat{\mu})U^{\dagger}_\mu(x+a\hat{\nu})U^{\dagger}_\nu(x),
\end{align}
where $U_{-\nu}(x)=U^{\dagger}_{\nu}(x-a\hat{\nu})$ and $P_{[\mu,\nu]}\equiv P_{\mu,\nu}-P_{\nu,\mu}$.

\subsection{Details of fits to eliminate the excited state contamination}

Using the power counting of $e^{-\delta m t_f}$, the summed three function can be rewritten into the following form,
\begin{eqnarray}
SC^{q_v,q_s,g}_3(t_f;H)&=&e^{-M_Ht_f}\left(B_0 t_f \langle O^{q_v,q_s,g}\rangle_H+B_4)+{\cal O}(e^{-\delta m t_f}\right), \nonumber\\
C_2(t_f;H)&=&e^{-M_Ht_f}B_0+{\cal O}(e^{-\delta m t_f})
\end{eqnarray}
And then one can obtain that $\langle O^{q_v,q_s,g}\rangle_H=\frac{SC^i_3(t_f;H)}{C_2(t_f;H)}-\frac{SC^i_3(t_f-1;H)}{C_2(t_f-1;H)}
+{\cal O}(e^{-\delta m t_f})$ in the large $t_f$ limit.

\begin{table}[ht!]
\caption{The $\chi^2$/d.o.f., $M_H$,  $\sum_q\langle m_q \bar{q}q\rangle_H$, -$\langle F^2\rangle_H$ and $\delta_m$ based on the joint fit for different channels with $m_v$=0.026, 0.092, 0.160, 0.319 and 0.479GeV.}
\begin{ruledtabular}
\begin{tabular}{c| ccc   ccc  ccc ccc ccc }
$m_v$(GeV) &  \multicolumn{3}{c}{0.026} &  \multicolumn{3}{c}{0.092}&  \multicolumn{3}{c}{0.160}  \\
Channel & PS & V & N  & PS & V & N  & PS & V & N   \\
\hline
$ \chi^2$/d.o.f. & 1.20 & 1.22 & 0.63 &1.18 & 0.68 & 0.84 &1.11 &0.74&0.82\\
$ M_H $(GeV) &0.340(1) & 0.881(4) &1.161(6) &0.647(1) &1.027(2) & 1.468(4) & 0.864(1)&1.174(2) &1.733(3)& \\
$\sum_q\langle m_q \bar{q}q\rangle_H$(GeV) &0.176(2) &0.126(5) &0.223(14) &0.341(2) &0.254(6) &0.447(10) &0.477(2) & 0.387(4) &0.645(9)\\
-$\langle F^2\rangle_H$(GeV) & 1.25(42) & 9.83(1.45)&10.92(1.48)&1.84(10)&8.84(45)&13.46(1.46)&2.23(12)&8.93(43)&11.77(1.07)\\
$ \delta m $(GeV) &1.07(12) & 1.22(17)& 0.72(7) & 1.00(5)&1.01(10)&0.61(4) &1.02(8)&0.82(5)&0.54(3) \\
\hline
$m_v$ & \multicolumn{3}{c}{0.319}&  \multicolumn{3}{c}{0.479}  \\
Channel  & PS & V & N  & PS & V & N \\
\hline
$ \chi^2$/d.o.f.  &1.11&1.06&0.98&0.96&1.26&1.18\\
$ M_H $(GeV) &1.277(1) &1.505(1) &2.280(2) &1.640(1) &1.825(1) &2.783(2)\\
$\sum_q\langle m_q \bar{q}q\rangle_H$(GeV) & 0.770(2)& 0.682(3) & 1.081(8)&1.047(1)&0.953(1) &1.478(1)\\
-$\langle F^2\rangle_H$(GeV) &2.77(11)&7.69(26) &10.31(90) &2.52(28) &6.52(60)&9.56(37)\\
$ \delta m $(GeV)  &0.78(4)&0.68(3)&0.48(2) & 0.72(1) & 0.66(4) &0.44(2)  \\
\end{tabular}
\end{ruledtabular}
\label{table:fitresult}
\end{table}

Using the joint fit of the summed 3pt and 2pt, we extract the matrix element of the ground state.
The fit results of $\chi^2$/d.o.f., mass of the ground state hadron $M_H$, the expectional vale of quark mass $\sum_q\langle m_q \bar{q}q\rangle_H$ and gluon operator -$\langle F^2\rangle_H$ and  $\delta_m$ are listed in Table ~\ref{table:fitresult} for all the hadrons used in this work.

\subsection{The trace anomaly density and radius in the hadrons}
We calculate the trace anomaly density using the following 
summed three point correlation function,
\begin{equation}\label{eq:fft}
SC_3^g(t_f,r;{\cal H})=\sum_{0<t<t_f}\sum_{\vec{x}}\sum_{|\vec{y}-\vec{x}|=r}\Big\langle\left[\langle C_2(t_f,\vec{x};{\cal H})-\langle C_2(t_f,\vec{x};{\cal H})\rangle\rangle\right][F^2(\vec{y},t)-\frac{1}{V}\langle \sum_{\vec{z}}F^2(\vec{z},t)\rangle]\Big\rangle,
\end{equation}
and then $\rho_H$ can be obtained by
\begin{equation}\label{eq:Rf}
\rho_H(r)=\frac{\beta}{2g\sum_{|\vec{y}-\vec{x}|=r}1}\big(\frac{SC_3^g(t_f,r;{\cal H})}{C_2(t_f;{\cal H})}-\frac{SC_3^g(t_f-1,r,{\cal H})}{C_2(t_f-1,\cal H)}\big)|_{t_f\rightarrow \infty, L\rightarrow \infty}.
\end{equation}
In practical, we use the joint fit defined in Eq. (8-9) to eliminate the excited state contamination with the data at finite $t_f$.

\begin{figure}[htbp]
	\centering
	\includegraphics[width=0.48\textwidth]{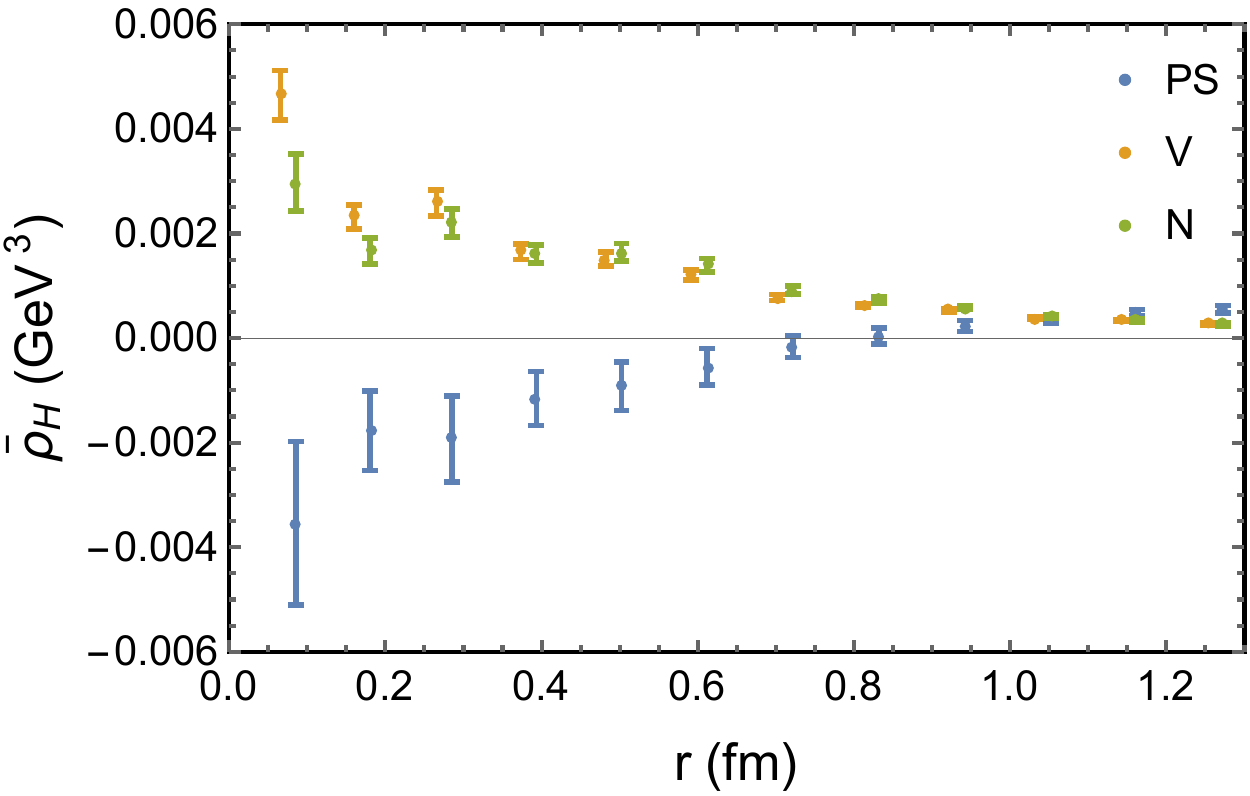}
	\caption{The normalized density of gluon trace anomaly in the different hadron state with 340 MeV pion mass.}
	\label{fig:radius}
\end{figure}

Generally, the charge radius can be defined through the integration of the normalized density $\bar{\rho}_H(r)=\frac{\rho_H(r)}{\int \textrm{d}^3r'' \rho_H(r'')}$,
\begin{equation}\label{eq:radius}
\langle r^2_g\rangle_H=\int \textrm{d}^3r' r'^2 \bar{\rho}_H(r').
\end{equation}
But in a finite volume $L^3\times T$, such a definition can only be accurate when $\bar{\rho}_H(r)$ is negligible at $r\sim L/2$, otherwise it will receive additional contribution from $\rho_H(L-r)$. Thus access the charge radius through the form factor $F_g(Q^2)$ as the function of the momentum transfer $Q^2$,
\begin{equation}
\langle r^2_g\rangle_H=\frac{6}{F_g(0)}\frac{\textrm{d}F_g(Q^2)}{\textrm{d}Q^2}|_{Q^2\rightarrow0},
\end{equation}
is a more widely used choice.

As shown in the Fig.~\ref{fig:radius}, $\bar{\rho}_H$ is still non-zero at $r=L/2$, especially for the pion case. Thus integrate it with $r^2$ to obtain the charge radius can have huge systematic uncertainty. But as in the figure, $\bar{\rho}_{\pi}(r)$ is smaller than $\bar{\rho}_{N}(r)$ for all the $r<$ 1.1 fm. Thus gerenrally we can rewrite the pion charge radius into
\begin{equation}\label{eq:radius2}
\langle r^2_g\rangle_{\pi}=\int \textrm{d}^3r' r'^2 \bar{\rho}_{\pi}(r')=\langle r^2_g\rangle_{N}+\int \textrm{d}^3r' (r'^2-r_0^2) (\bar{\rho}_{\pi}(r')-\bar{\rho}_{N}(r')),
\end{equation}
where $r_0$ is the intersection of $\bar{\rho}_{\pi}(r_0)$ and $\bar{\rho}_{N}(r_0)$ and we have $\bar{\rho}_{\pi}(r_0)=\bar{\rho}_{N}(r_0)$ at $r_0$. Then the integrand of the second term in the right hand side will be always positive, and one can expect $\langle r^2_g\rangle_{\pi}>\langle r^2_g\rangle_{N}$.
 
\end{widetext}
\end{document}